# Room temperature observation of biexcitons in exfoliated WS$_2$ monolayers

I. Paradisanos[1,2], S. Germanis[1], N. T. Pelekanos[1,3], C. Fotakis[1,2], E. Kymakis[1,4], G. Kioseoglou*[1,3], and E. Stratakis*[1,3]

[1] *Institute of Electronic Structure and Laser, Foundation for Research and Technology - Hellas, Heraklion, 71110, Crete, Greece*

[2] *Department of Physics, University of Crete, Heraklion, 71003, Crete, Greece*

[3] *Department of Materials Science and Technology, University of Crete, Heraklion, 71003 Crete, Greece*

[4] *Center of Materials Technology and Photonics & Electrical Engineering Department, Technological Educational Institute (TEI) of Crete, Heraklion, 71004 Crete, Greece*

**Abstract**

Single layers of WS$_2$ are direct gap semiconductors with high photoluminescence (PL) yield holding great promise for emerging applications in optoelectronics. The spatial confinement in a 2D monolayer together with the weak dielectric screening lead to huge binding energies for the neutral excitons as well as other excitonic complexes, such as trions and biexcitons whose binding energies scale accordingly. Here, we report on the existence of biexcitons in mechanically exfoliated WS$_2$ flakes from 78 K up to room temperature. Performing temperature and power dependent PL measurements, we identify the biexciton emission channel through the superlinear behavior of the integrated PL intensity as a function of the excitation power density. On the contrary, neutral and charged excitons show a linear to sublinear dependence in the whole temperature range. From the energy difference between the emission channels of the biexciton and neutral exciton, a biexciton binding energy of 65-70 meV is determined.

*Corresponding authors: gnk@materials.uoc.gr ; stratak@iesl.forth.gr

The interest on two-dimensional (2D) materials has been steadily increasing since the discovery of graphene, a material with fascinating properties and great potential for various applications[1]. Transition metal dichalcogenides (TMDs) with the form $MX_2$ (M = Mo, W, Ti, etc and X = S, Se, Te) exhibit a structure very similar to that of graphene and have attracted significant attention of the scientific community [2-5] for a number of reasons. Among the most attractive features of TMD compounds is that their electronic properties can vary from metallic to those of a wide band gap semiconductor depending on the structure, composition and dimensionality, while their band structure can be easily tuned by applying stress[6]. Owing to such layer-dependent electronic structure, TMDs exhibit extraordinary optoelectronic properties[7,8] as well as potential for enhanced performance in applications such as photodetectors[9], photovoltaics[10,11] and non-linear optical components[12].

Among their prominent optical properties, TMD monolayers show strong PL in the visible and near-infrared spectral range, due to the transition from indirect band gap semiconductors in their bulk and few-layer forms, to direct band gap semiconductors in the monolayer limit[13]. Besides this, the spatial confinement of carriers in a 2D monolayer lattice and the weak dielectric screening, give rise to unusually strong excitonic effects and high binding energies[14-16]. These properties favor the stability of a variety of excitonic quasiparticles with extremely large binding energies, including neutral excitons with binding energies of several hundred meV [17,18], charged excitons (trions)[19,20] and exciton-trion complexes[21] exhibiting binding energies of tens of meV.

Besides excitons and trions, ultrathin TMDs can additionally host bound quasiparticles that individually consist of two electron-hole pairs, known as biexcitons. The presence of biexcitons is exciting due to their unique role in understanding many body effects and their great promise for practical applications[22]. Apart from the large binding energies, it is also expected to show unique properties including entanglement between the pair of valley pseudospins[16]. In this context, thorough investigation of biexcitons is crucial to assess the potential of TMDs in optoelectronic applications. Although trions in monolayer TMDs are relatively well understood[20,23,24] the existence of biexcitons has only been recently reported at cryogenic temperatures in $WS_2$,[25] and $WSe_2$, [26] exfoliated monolayers. More importantly, it was shown that the biexciton emission in $WS_2$ can be electrically tuned[27], which is of great importance for TMDs-based photonic devices.

The unusually strong excitonic effects in TMDs strongly suggest an enhanced thermal stability of biexciton emission. While neutral and charged excitons have been extensively studied, the direct observation of biexciton emission at elevated temperatures is still quite unexplored. Indeed, there is currently no report on the existence of biexcitons at temperatures higher than 60K in TMD monolayers, exfoliated from the bulk crystal. Such investigation would provide important insight into the physics of quantum confined excitonic effects in TMDs, considering that it is quite challenging to identify distinct excitonic features at elevated temperatures where exciton - phonon interaction is quite strong. In this letter, we investigate the excitonic features of mechanically exfoliated $WS_2$ monolayers, via the excitation power dependence of the PL intensity in a temperature range from 78K to 295K. Based on the analysis of the different excitonic peaks, we postulate the presence of non-linear excitonic features in the form of biexcitons up to room temperature.

WS$_2$ samples were mechanically exfoliated from a bulk crystal (2D semiconductors) using an adhesive tape and were subsequently deposited on Si/Silicon Oxide (290nm) wafers. Monolayer regions of 5–8 µm across were identified via optical microscopy and confirmed with micro-Raman (Thermo Scientific) spectroscopy with very low laser power in order to avoid structural damage. The energy difference between the two main Raman vibrational modes, E′ (in plane mode) and A$_1$′ (out of plane mode), is used extensively in the literature as the fingerprint of the number of layers. Specifically, energy differences of 59-61 cm$^{-1}$ definitely confirm the existence of WS$_2$ monolayers[28]. Micro-photoluminescence (µ-PL) spectra were collected in a backscattering setup using an excitation source of 532 nm continuous wave laser through a 40x objective with numerical aperture of 0.65 forming a 1 µm spot diameter. Studies from 78 K to 295 K were performed in a liquid N$_2$ (LN) flow cryostat. The excitation power was controlled with a combination of neutral density filters. A spatial filter setup was used to obtain the central part of the Gaussian beam and enhance uniformity of the excitation spot.

Figure 1(a) shows a typical optical microscopy image of bulk and monolayer WS$_2$ flakes. A corresponding Raman spectrum obtained from the monolayer area is depicted in Figure 1(b). The energy difference between the E′ and A$_1$′ vibrational modes stays always lower than 60 cm$^{-1}$, which unambiguously confirms the monolayer existence[28].

The excitation power dependence of the PL intensity can provide insight into the physical origin of the corresponding PL peaks. Specifically, high-power excitation can also lead to the emergence of additional PL peaks originating from more complex excitonic states such as biexcitons. Figure 2(a) shows the PL spectra of monolayer WS$_2$ at 78 K with different excitation powers. For the analysis of the various emission peaks, we fit the spectra with Gaussian functions achieving very high values of coefficient of determination ($R^2$>0.99). It can be observed that the linewidths of the different emission peaks at 78K are sufficiently small, giving rise to well-resolved features in the spectra. For each power, the corresponding PL spectra are deconvoluted into four peaks (inset Figure 2(a)), which are assigned (from higher to lower energy) to neutral exciton (X), negatively charged exciton (trion, $X^-$), superposition of biexciton (XX) and defect-bound exciton (D) (XX/D) and emission from localized states (L)[25]. The L peak probably originates from disorder effects[29] and impurities[27]. Although at low excitation powers, $X^-$ and XX/D are spectrally well separated and of comparable intensity, the XX/D peak dominates the spectrum at higher excitation powers. To get an understanding of the nature of the XX/D peak, we monitor the integrated PL intensity of the main emission channels (L, XX/D, $X^-$ and X) as a function of the excitation density, displayed in the double-logarithmic plot in Figure 2(b). In accordance with previous studies[25,27] the L emission band exhibits a sublinear behavior[17,19]. The neutral exciton and the trion show a linear behavior which is expected for excitonic transitions.[15] On the contrary, at low excitation power densities the XX/D peak intensity exhibits a linear dependence, while for densities higher than ~1 kW/cm$^2$, it exhibits a superlinear relationship with a power law exponent of 1.4. The emission strength of the XX/D peak at different excitation densities is also depicted in the 3D plot of Figure (S1). At thermal equilibrium (ignoring possible temperature increase at elevated excitation densities), we would expect a quadratic relation between the biexciton PL integrated intensity and the excitation power density[30], that is $I_{PL} \propto P_{ex}^2$. However, exponents of 1.2-1.9 are commonly observed in quantum-well systems and attributed to the lack of equilibrium between the states[31], as well as to the kinetics of biexciton formation and exciton recombination[26]. The difference between low and high excitation densities strongly suggest that two different emission channels are responsible for the XX/D peak: at low excitation densities, the main contribution originates from defect-bound excitons (D), while at high excitation densities, the biexciton (XX) emission becomes prominent.

Figure 2(c) shows the peak positions of L, X, X⁻, and XX/D, from which we estimate the binding energies of trions and biexcitons to be 40 meV and 65 meV, respectively. These values are consistent with previous observations on mechanically exfoliated monolayer WS$_2$[25]. However, it should be noted that there is significant difference in trion binding energy between CVD as-grown[32] (i.e ~20 meV) and our mechanically exfoliated monolayer WS$_2$ (i.e ~40 meV). This is possibly due to the different doping levels and strain within the as-grown sample[20]. It is also evident from Figure 2(c) that the XX/D peak redshifts by approximately 5 meV in the excitation density range studied. This is an indication that the D emission from defect-bound excitons at low excitation densities is at a slightly higher energy than the biexciton emission at high excitation densities. Since neither X or X⁻ peak displays a redshift, the observed small redshift of the 5 meV for the XX/D peak cannot be attributed to local heating induced by the excitation source. This small redshift was taken into account for the estimation of the biexciton binding energy of 65 meV. The binding energy of the biexciton ($E_{bXX}$) is defined as the energy difference between two free excitons and the biexciton energy state, $E_{bXX} = 2E_X - E_{XX}$, where $E_X = \hbar\omega_X$ is the energy of the 1s exciton state determined by the corresponding PL emission energy, and $E_{XX}$ is the energy that corresponds to the biexciton state. $E_{XX}$ can be determined from the biexciton PL at energy $\hbar\omega_{XX}$ and by assuming that radiative decay of the biexciton produces an exciton, then $E_{XX} = \hbar\omega_{XX} + \hbar\omega_X$. Therefore, the biexciton binding energy is the energy difference between the emission channels of the neutral exciton and the biexciton $E_{bXX} = 2\hbar\omega_X - (\hbar\omega_{XX} + \hbar\omega_X) = \hbar\omega_X - \hbar\omega_{XX}$

. From Figure 2(c) this energy difference is 65 meV taking into account the small red shift of the XX/D peak after 2 kW/cm$^2$. Recent theoretical calculations accurately predict the binding energies of excitons and trions within 2D materials, but they fail to do so for biexcitons[33-39]. The models predict that the biexciton is less strongly bound than the trion, in contradiction to experimental findings [25,27]. However, the models reconcile this discrepancy by suggesting that the experimentally observed states may actually be those of excited biexcitons, as opposed to ground state biexcitons[25]. This suggestion is very controversial because it is difficult to explain the absence of the ground state of an excitonic complex. Other calculations give a good estimate of the binding energy of biexcitons considering the electron at K-valley to be bound to a hole at K' and vice versa or assuming a strictly local screening[25]. Another plausible explanation could be that in the experiment the biexcitons are bound to, e.g., impurities or defects and that is why the slope is less than two as one would expect from biexcitons. Therefore, further experimental and theoretical work is needed to fully explain this discrepancy. Our observations above are consistent with recent works on low-temperature PL features of mechanically exfoliated monolayer and as-grown CVD bilayer WS$_2$. Specifically Plechinger et al.[25] reported on the emergence of the XX/D peak below 60 K for monolayer WS$_2$, while He et al.[40] reported on biexcitons in bilayer WS$_2$ at 77 K with difficulties resolving the biexciton emission at temperatures higher than 150K.

To provide a further insight into the stability of the observed high-order excitonic effects and their thermal properties, power-dependent PL spectra have been recorded at elevated temperatures. Figure 3(a) presents the PL spectra under different excitation powers, for the same WS$_2$ monolayer, at T=200 K. Interestingly, for excitation power densities higher than 0.7kW/cm$^2$ the main PL emission located at ~2.00 eV cannot be fitted with a single peak (red arrow in Figure 3(a), X$^-$) exhibiting a noticeable asymmetry at lower energies, indicating a definite contribution from an additional lower energy peak located at 1.985 eV which we attribute to biexciton emission (blue arrow in Figure 3(a), XX). The logarithmic scale of the y-axis in Figure 3(a) further highlights the aforementioned asymmetry. As expected, the trion and biexciton emission peaks are not well-resolved in the 200K PL spectrum since their linewidths strongly increase from ~17meV at 78K to almost 40meV at 200K, due to phonon-related thermal broadening effects. In view of the above observations, the PL spectra were deconvoluted into three peaks, assigned as X, X$^-$ and XX, as shown in the inset of Figure 3(a). While the integrated intensities of X and X$^-$ show a roughly linear dependence on the excitation power (Figure 3(b)), the XX peak exhibits a non-linear dependence in the whole range of excitation densities with a power law exponent of 1.3, close to the value of 1.4 measured at 78K. However, it is important to note that when considering the ratio between the intensity of biexcitons ($I_{XX}$) and of neutral excitons ($I_X$), a power law exponent of 1.7 is obtained (Figure 3(d)). Such superlinear dependence further confirms the presence of biexciton state.[18] From Figure 3(b), it is also clear that there is no contribution from the defect related component since the XX data exhibit only one slope for all excitation powers. In addition, the peak position remained constant with increasing excitation density, revealing a binding energy of ~70meV, excluding the slight red shift for values higher than 50kW/cm$^2$ which can be attributed to thermal effects (Figure 3(c)).

Surprisingly, the biexcitonic features observed at 200K persist even under ambient conditions. This is evident in the power dependence of the integrated PL intensity at T=295 K, shown in Figure 4(a). At room temperature the neutral exciton is observed only at low and medium excitation powers. By increasing the excitation power, the linewidths of the X$^-$ and XX overlap with the weak X emission; as a result, the fitting of the exciton for high excitation powers is not reliable. However, for excitation densities where the fitting of the neutral exciton is valid a linear dependence on the excitation density is observed (Figure 4(b)). We will focus our discussion now on the power dependence of the trion and biexciton lines. While the trion shows an almost sublinear dependence on the excitation density, the XX peak exhibits a superlinear dependence with a power law exponent of 1.3 (Figure 4(b)). It is important to note that similar to the case of 200K, the main emission peak cannot be fitted with a single Gaussian function for excitation powers higher than 0.7kW/cm$^2$, due to a pronounced asymmetry on the low energy side. The redshift of both X$^-$ and XX peaks in Figure 4(c) for powers beyond 50kW/cm$^2$ is due to thermal effects. However, possible relation between the PL spectrum asymmetry and laser induced heating can be ruled out since the peak positions remained constant for excitation powers at which spectral asymmetry appeared. From the biexciton position in the power range before heating effects take place, a biexciton binding energy of ~70meV can be determined. Such a predominant biexciton emission has not yet been observed from an exfoliated TMD monolayer at room temperature. It should be mentioned here that the temperature dependence of the neutral exciton, trion and biexciton, shown in Figure (S2), exhibit a standard excitonic behavior and can be described well by the Varshni relation with parameters $α=0.6$ meV/K and $β=280$ K. In addition, the temperature-dependent PL spectra under 3.8kW/cm$^2$ excitation power density is presented in Figure (S3).

Further experimental investigations are currently in progress in order to elucidate the nature of room temperature biexcitons in WS$_2$, based on time-resolved fluorescence spectroscopy complemented with selective polarization studies. Nevertheless, the results presented here prove that under suitable conditions the biexciton channel can be the strongest emission mechanism for a monolayer WS$_2$, even at room temperature. This four-body correlated system in ambient conditions reflects the unusual strength and thermal stability of many-body interactions in atomically thin TMDs. Besides this, the degeneracy of K and K´ valleys, enables the possible creation of optically bright biexciton states with pairs of carriers at different valleys. This not only generates uncommon types of quantum coherent excitations, but also opens the way for the creation of correlated photon pairs for the control of polarized states[41].

In conclusion, the observation of biexcitons in a temperature range from 78K up to room temperature on mechanically exfoliated $WS_2$ monolayers, has been demonstrated. Quantitative analysis of the PL spectra allowed us to identify an emission peak exhibiting a superlinear relationship between the integrated emission intensity and the excitation power density, with a power law exponent of 1.4 at 78K and 1.3 at 200K and 295K. Based on the analysis of the excitonic peaks emission energy, we deduced a biexciton binding energy of 65meV at 78K and 70meV at 200K and 295K. We envisage that our observations of the biexciton emission from exfoliated $WS_2$ monolayers in such a broad range of temperatures will be useful towards practical applications, such as biexciton lasing and in understanding many body effects in quantum confined systems.

## ACKNOWLEDGMENTS

This work is supported by the European Research Infrastructure NFFA-Europe, funded by EU's H2020 framework program for research and innovation under grant agreement n. 654360.

**FIGURES**

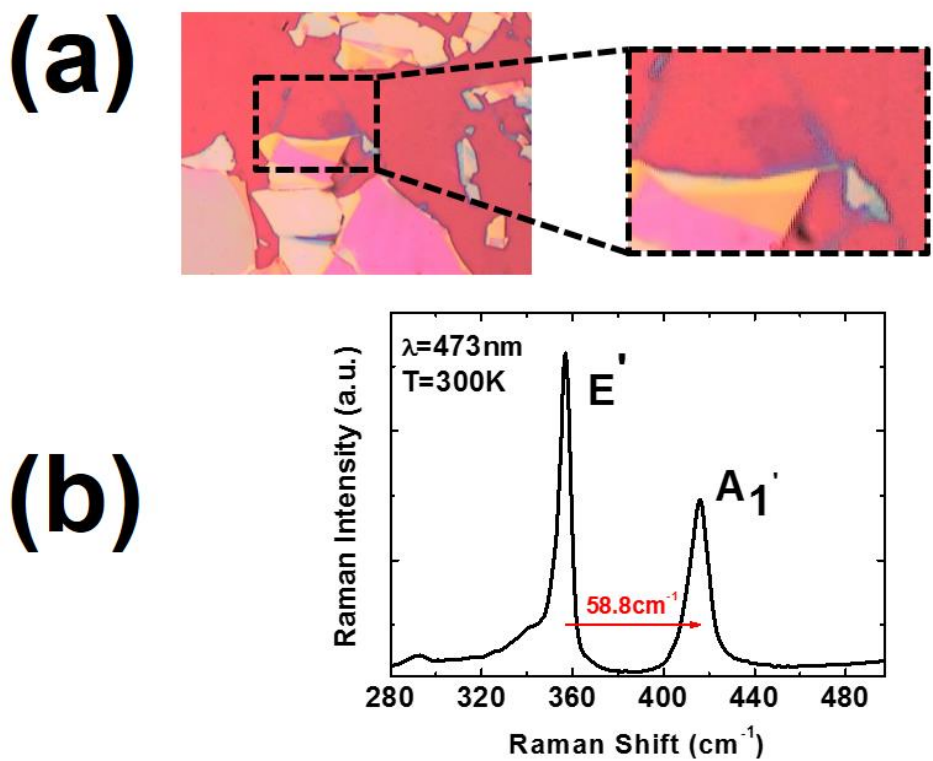

**Figure 1:** (a) Optical microscopy image of bulk and monolayer WS$_2$ flakes. (b) Raman spectrum obtained from the monolayer area. The 58.8cm$^{-1}$ energy difference between the in plane E´ and the out of plane A$_1$´ vibrational modes confirms the monolayer existence.

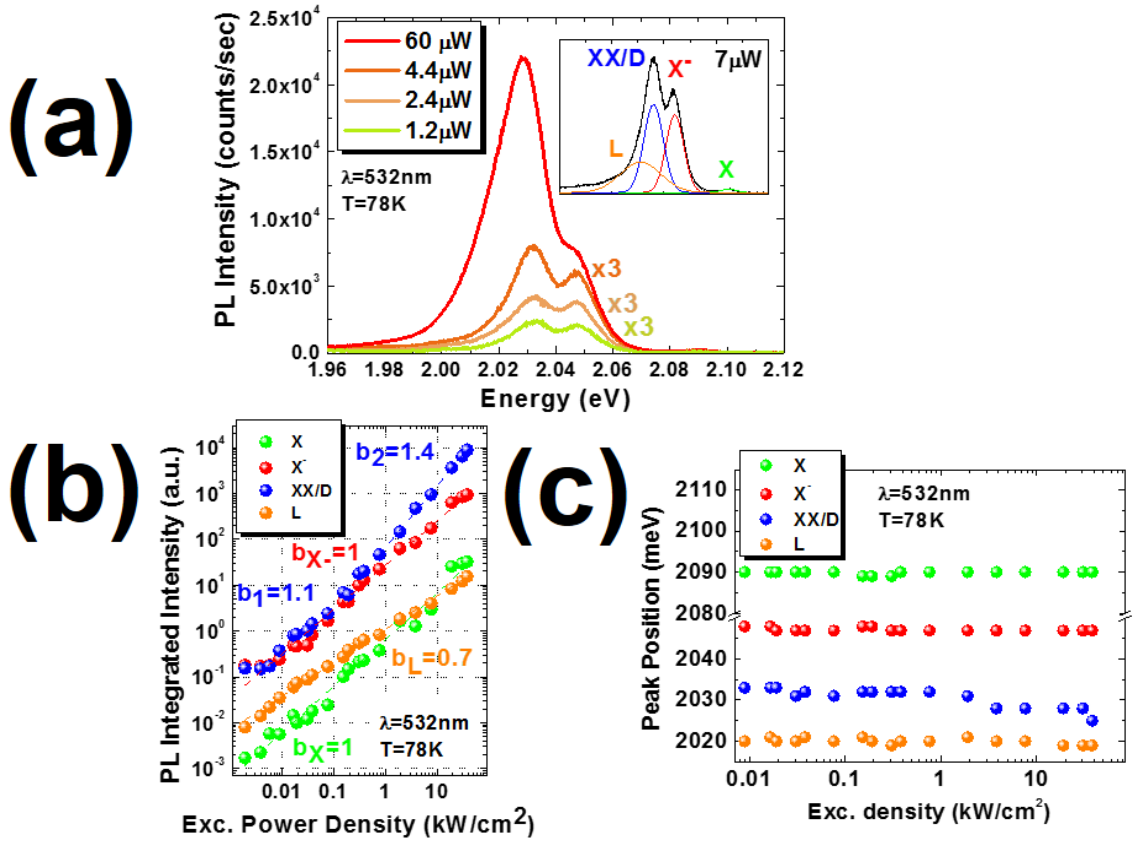

**Figure 2:** (a) PL spectra of monolayer $WS_2$ at 78 K with different excitation powers. Inset: deconvolution of PL spectrum taken at 78K with 7µW excitation power. Neutral exciton (X), negatively charged exciton (trion, $X^-$), superposition of defect-bound exciton (D) and biexciton (XX), (XX/D) and emission from localized states (L) are assigned. (b) Integrated PL intensity of L, XX/D, $X^-$ and X peaks for different excitation densities. The XX/D peak exhibits a superlinear relationship with a power law exponent of 1.4 for excitation densities higher than ~1kW/cm². (c) Peak positions of X, $X^-$, XX/D and L at 78K.

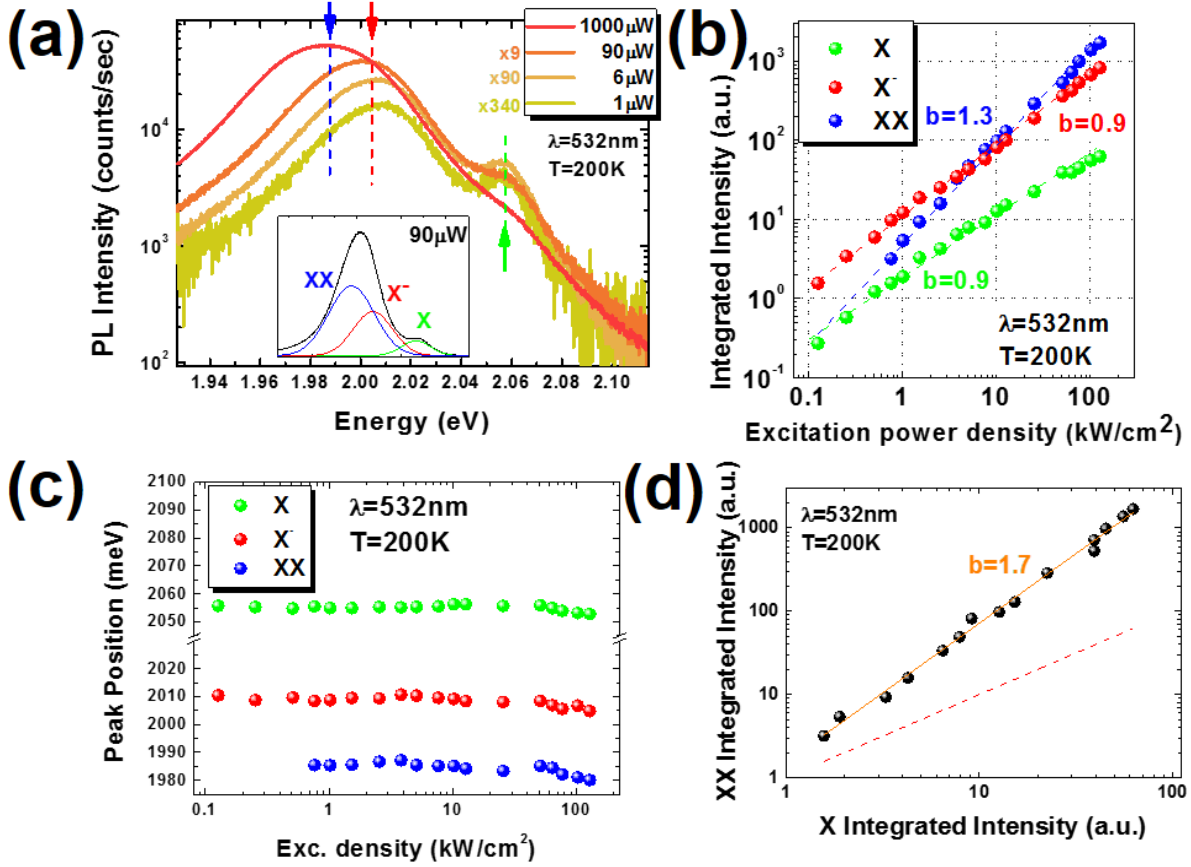

**Figure 3:** (a) PL spectra of monolayer $WS_2$ under different excitation powers, at T=200 K. Green, red and blue arrows indicate the position of neutral exciton (X), charged exciton ($X^-$) and biexciton (XX), respectively. Inset: deconvolution of PL spectrum taken at 200K with 90μW excitation power. Neutral exciton (X), negatively charged exciton (trion, $X^-$) and biexciton (XX) peaks are assigned. (b) Integrated PL intensity for XX, $X^-$ and X peaks for different excitation densities at 200K. The XX peak exhibits a superlinear relationship with a power law exponent of 1.3. (c) Peak positions of X, $X^-$, and XX at 200K. (d) Logarithmic plot of the biexciton integrated intensity (XX), as a function of the exciton integrated intensity (X). The orange line is a power-law fit: Integrated $I_{XX}$ / (Integrated $I_X$)$^b$, with b=1.7. For comparison, a linear relation (b=1) is shown as a dashed red line.

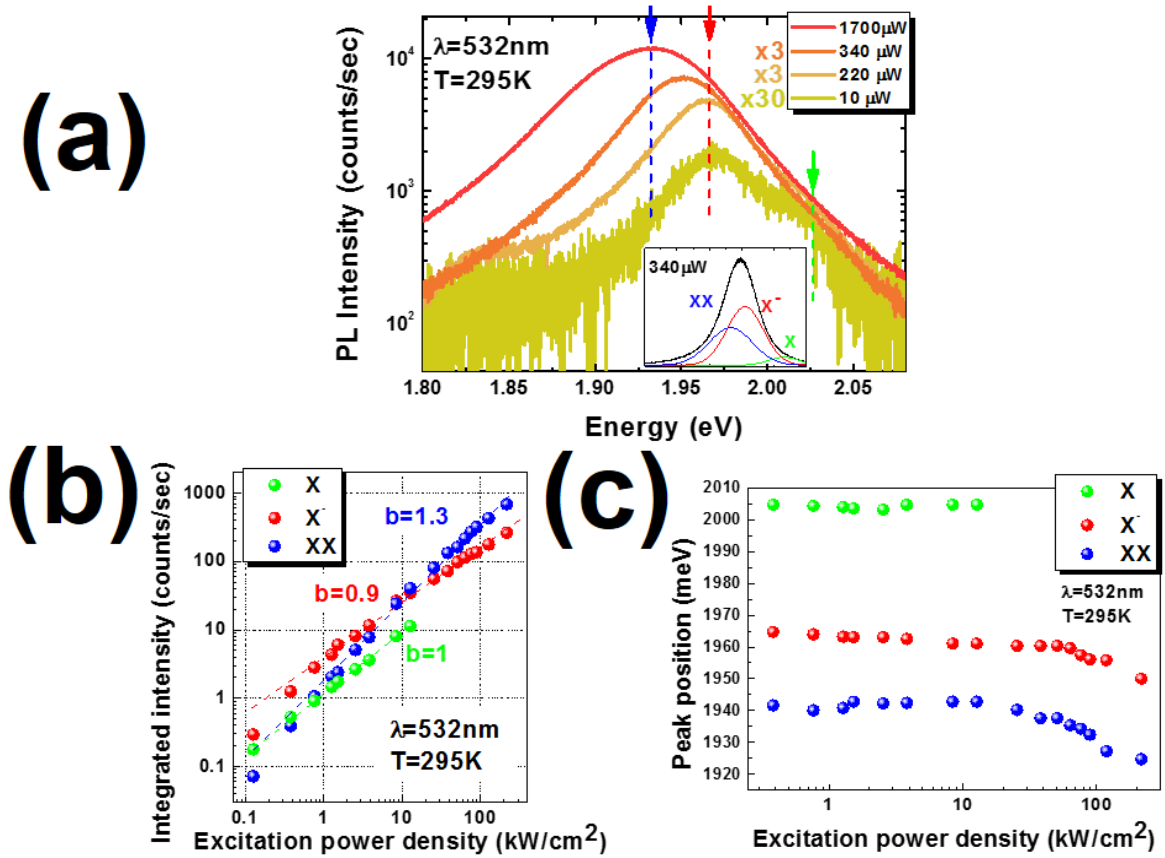

**Figure 4:** (a) PL spectra of monolayer $WS_2$ under different excitation powers, at T=295 K. Green, red and blue arrows indicate the position of neutral exciton (X), charged exciton($X^-$) and biexciton (XX), respectively. Inset: deconvolution of PL spectrum taken at 295K with 340μW excitation power. Neutral exciton (X), negatively charged exciton (trion, $X^-$) and biexciton (XX) are assigned. (b) Integrated PL intensity for XX $X^-$ and X for different excitation densities, at 295K. The XX peak exhibits a superlinear relationship with a power law exponent of 1.3. (c) Peak positions of XX $X^-$ and X at 295K. The redshift of both the $X^-$ and XX peaks for powers beyond 50kW/$cm^2$ is due to thermal effects.

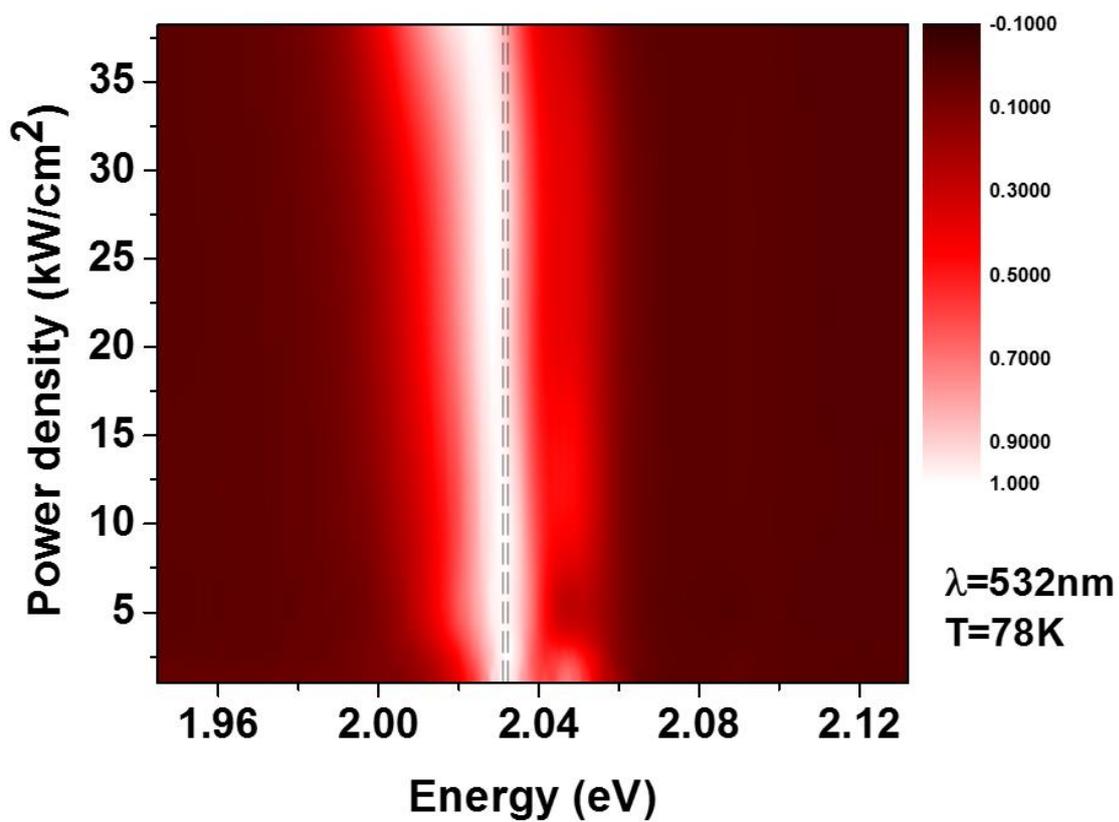

**Figure S1:** 3D plot showing the emission strength of the XX/D at different excitation densities, at 78K. With increasing the excitation power density the biexciton emission dominates the spectra.

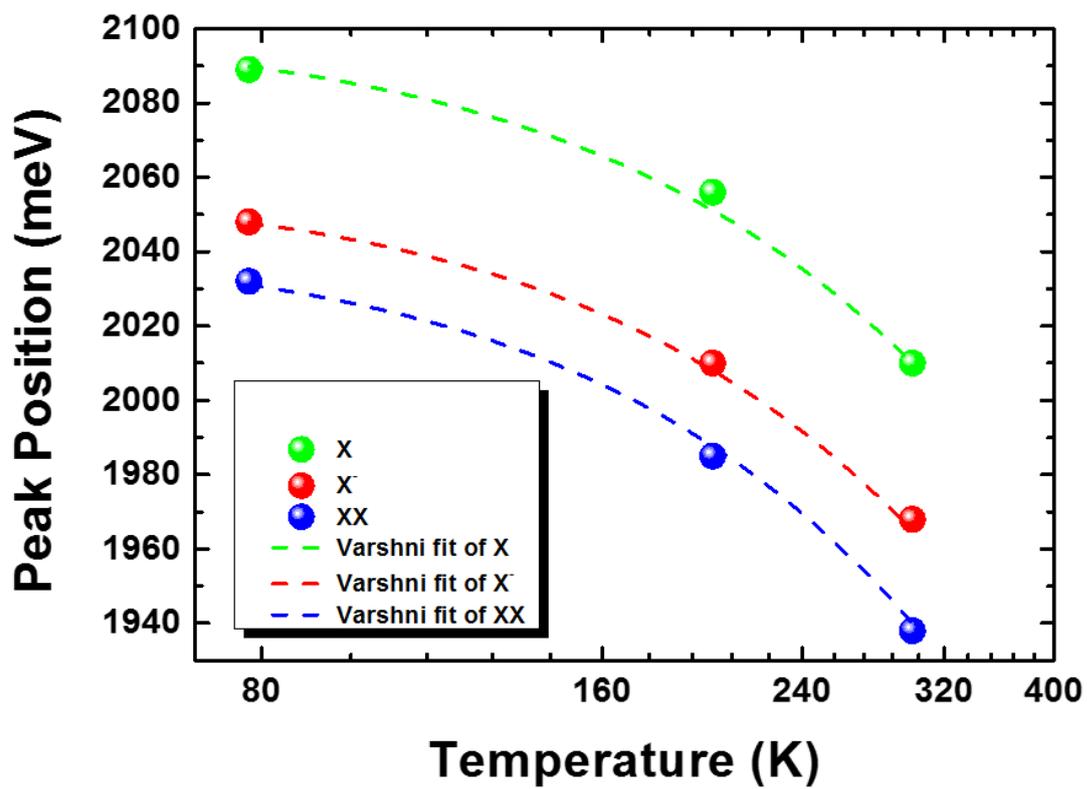

**Figure S2:** Temperature dependence of the neutral exciton, trion and biexciton exhibiting a typical excitonic behavior.

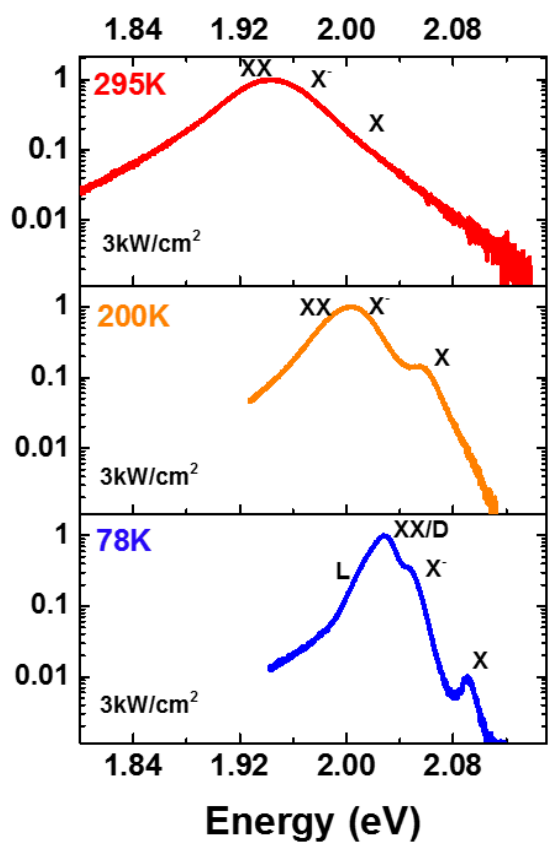

**Figure S3:** Temperature-dependent PL spectra under 3.8kW/cm$^2$ excitation power density.

**REFERENCES**

[1] K. S. Novoselov, A. K. Geim, S. V. Morozov, D. Jiang, Y. Zhang, S. V. Dubonos, I. V. Grigorieva, and A. A. Firsov, Science **306**, 666 (2004).

[2] Y. Lin and J. W. Connell, Nanoscale **4**, 6908 (2012).

[3] Y. Kubota, K. Watanabe, O. Tsuda, and T. Taniguchi, Science **317**, 932 (2007).

[4] C. R. Dean, A. F. Young, I. Meric, C. Lee, L. Wang, S. Sorgenfrei, K. Watanabe, T. Taniguchi, P. Kim, K. L. Shepard, and J. Hone, Nat Nano **5**, 722 (2010).

[5] Q. H. Wang, K. Kalantar-Zadeh, A. Kis, J. N. Coleman, and M. S. Strano, Nat Nano **7**, 699 (2012).

[6]   M. Chhowalla, H. S. Shin, G. Eda, L.-J. Li, K. P. Loh, and H. Zhang, Nat Chem **5**, 263 (2013).

[7]   K. F. Mak, C. Lee, J. Hone, J. Shan, and T. F. Heinz, Physical Review Letters **105**, 136805 (2010).

[8]   B. Radisavljevic, A. Radenovic, J. Brivio, V. Giacometti, and A. Kis, Nat Nano **6**, 147 (2011).

[9]   O. Lopez-Sanchez, D. Lembke, M. Kayci, A. Radenovic, and A. Kis, Nat Nano **8**, 497 (2013).

[10]  M. M. Furchi, A. Pospischil, F. Libisch, J. Burgdörfer, and T. Mueller, Nano Letters **14**, 4785 (2014).

[11]  M. Sygletou, P. Tzourmpakis, C. Petridis, D. Konios, C. Fotakis, E. Kymakis, and E. Stratakis, Journal of Materials Chemistry A **4**, 1020 (2016).

[12]  C. Janisch, Y. Wang, D. Ma, N. Mehta, A. L. Elías, N. Perea-López, M. Terrones, V. Crespi, and Z. Liu, Scientific Reports **4**, 5530 (2014).

[13]  A. Splendiani, L. Sun, Y. Zhang, T. Li, J. Kim, C.-Y. Chim, G. Galli, and F. Wang, Nano Letters **10**, 1271 (2010).

[14]  G. Berghäuser and E. Malic, Physical Review B **89**, 125309 (2014).

[15]  Z. Ye, T. Cao, K. O'Brien, H. Zhu, X. Yin, Y. Wang, S. G. Louie, and X. Zhang, Nature **513**, 214 (2014).

[16]  E. J. Sie, A. J. Frenzel, Y.-H. Lee, J. Kong, and N. Gedik, Physical Review B **92**, 125417 (2015).

17    M. M. Ugeda, A. J. Bradley, S.F. Shi, F. H. da Jornada, Y. Zhang, D. Y. Qiu, W. Ruan, S. K. Mo, Z. Hussain, Z. X. Shen, F. Wang, S. G. Louie, and F. Crommie, Nat Mater **13**, 1091 (2014).

18    G. Wang, X. Marie, I. Gerber, T. Amand, D. Lagarde, L. Bouet, M. Vidal, A. Balocchi, and B. Urbaszek, Physical Review Letters **114**, 097403 (2015).

19    H. M. Hill, A. F. Rigosi, C. Roquelet, A. Chernikov, T. C. Berkelbach, D. R. Reichman, M. S. Hybertsen, L. E. Brus, and T. F. Heinz, Nano Letters **15**, 2992 (2015).

20    K. F. Mak, K. He, C. Lee, G. H. Lee, J. Hone, T. F. Heinz, and J. Shan, Nat Mater **12**, 207 (2013).

21    A. Singh, G. Moody, S. Wu, Y. Wu, N. J. Ghimire, J. Yan, D. G. Mandrus, X. Xu, and X. Li, Physical Review Letters **112**, 216804 (2014).

22    J. Q. Grim, S. Christodoulou, F. Di Stasio, R. Krahne, R. Cingolani, L. Manna, and I. Moreels, Nat Nano **9**, 891 (2014).

23    M. Currie, A. T. Hanbicki, G. Kioseoglou, and B. T. Jonker, Applied Physics Letters **106**, 201907 (2015).

24    I. Paradisanos, N. Pliatsikas, P. Patsalas, C. Fotakis, E. Kymakis, G. Kioseoglou, and E. Stratakis, Nanoscale **8**, 16197 (2016).

25    G. Plechinger, P. Nagler, J. Kraus, N. Paradiso, C. Strunk, C. Schüller, and T. Korn, physica status solidi (RRL) – Rapid Research Letters **9**, 457 (2015).

26    Y. You, X.-X. Zhang, T. C. Berkelbach, M. S. Hybertsen, D. R. Reichman, and T. F. Heinz, Nat Phys **11**, 477 (2015).

[27] J. Shang, X. Shen, C. Cong, N. Peimyoo, B. Cao, M. Eginligil, and T. Yu, ACS Nano **9**, 647 (2015).

[28] W. Zhao, Z. Ghorannevis, K. K. Amara, J. R. Pang, M. Toh, X. Zhang, C. Kloc, P. H. Tan, and G. Eda, Nanoscale **5**, 9677 (2013).

[29] B. Bansal, A. Kadir, A. Bhattacharya, and V. V. Moshchalkov, Applied Physics Letters **93**, 021113 (2008).

[30] J. C. Kim, D. R. Wake, and J. P. Wolfe, Physical Review B **50**, 15099 (1994).

[31] R. T. Phillips, D. J. Lovering, G. J. Denton, and G. W. Smith, Physical Review B **45**, 4308 (1992).

[32] M. S. Kim, S. J. Yun, Y. Lee, C. Seo, G. H. Han, K. K. Kim, Y. H. Lee, and J. Kim, ACS Nano **10**, 2399 (2016).

[33] D. K. Zhang, D. W. Kidd, and K. Varga, Nano Letters **15**, 7002 (2015).

[34] D. W. Kidd, D. K. Zhang, and K. Varga, Physical Review B **93**, 125423 (2016).

[35] M. Z. Mayers, T. C. Berkelbach, M. S. Hybertsen, and D. R. Reichman, Physical Review B **92**, 161404 (2015).

[36] R. Y. Kezerashvili and S. M. Tsiklauri, Few-Body Systems **58**, 18 (2016).

[37] K. A. Velizhanin and A. Saxena, Physical Review B **92**, 195305 (2015).

[38] A. Thilagam, Journal of Applied Physics **116**, 053523 (2014).

[39] I. Kylänpää and H.-P. Komsa, Physical Review B **92**, 205418 (2015).

[40] Z. He, W. Xu, Y. Zhou, X. Wang, Y. Sheng, Y. Rong, S. Guo, J. Zhang, J. M. Smith, and J. H. Warner, ACS Nano **10**, 2176 (2016).

[41] N. Akopian, N. H. Lindner, E. Poem, Y. Berlatzky, J. Avron, D. Gershoni, B. D. Gerardot, and P. M. Petroff, Physical Review Letters **96**, 130501 (2006).